\documentclass[runningheads]{llncs}
\usepackage{graphicx}
\usepackage{tabularx}
\usepackage{array}
\usepackage{multirow}
\usepackage{arydshln}
\usepackage{dblfloatfix}
\usepackage{tikz}
\usepackage{calc}
\usepackage{amsmath}
\usepackage{mathtools}
\usepackage{amsfonts}

\begin{document}
\title{OAK: Ontology-based Knowledge Map\\
       Model for Digital Agriculture}
%
\author{Quoc Hung Ngo \and Tahar Kechadi \and Nhien-An Le-Khac}
%

%
\institute{School of Computer Science, College of Science\\
University College Dublin, Belfield, Dublin 4, Ireland\\
\email{hung.ngo@ucdconnect.ie},
\email{\{tahar.kechadi, an.lekhac\}@ucd.ie}}
\maketitle             
\begin{abstract}

  Nowadays, a  huge amount  of knowledge  has been amassed  in digital  agriculture.  This
  knowledge and know-how information are collected from various sources, hence the question
  is how  to organise this  knowledge so that it  can be efficiently  exploited.  Although
  this knowledge about agriculture practices can be represented using ontology, rule-based
  expert systems,  or knowledge model  built from  data mining processes,  the scalability
  still  remains an  open issue.   In this  study, we  propose a  knowledge representation
  model,  called  an  ontology-based  knowledge  map, which  can  collect  knowledge  from
  different sources, store it, and exploit either  directly by stakeholders or as an input
  to the  knowledge discovery process (Data  Mining).  The proposed model  consists of two
  stages, 1) build an  ontology as a knowledge base for a specific  domain and data mining
  concepts,  and 2)  build the  ontology-based knowledge  map model  for representing  and
  storing the knowledge mined on the crop  datasets. A framework of the proposed model has
  been implemented in agriculture  domain. It is an efficient and scalable  model, and  it 
  can be used as knowledge repository a digital agriculture.
.

\end{abstract}

\section{Introduction}
\label{sec:intro}

The knowledge  from crop studies  is turned into  profitable decisions in  digital farming
only when it is efficiently managed. Farming knowledge can be created by the experience of
farmers, by research studies  in agronomy, or by analyses of data  that has been collected
for a number of years. In particular,  the knowledge discovery in agricultural data is one
of the most diverse, large, and dynamic in digital farming.

Moreover, one of the  key challenges of knowledge management is  the representation of the
mined  results, which  are  not  consistent among  different  sources  of knowledge.   For
example, consider two  results that were obtained  from two data mining  studies, with the
overall goal  of predicting farming  conditions of high crop  yield of winter  wheat.  The
concept \textit{high crop  yield} used in the  two studies can be  different, depending on
the  way authors  define the  range of  \textit{high crop  yield}. Therefore,  the overall
insight  may not  be  consistent  to be  used  in  the same  system  or  to compare  those
results. This  issue can  also occur  with input attributes  used by  different predictive
models. In addition, the farming knowledge represented  using an ontology is static and it
is  difficult   to  generalise  to   different  regions,  which  have   different  farming
conditions. The knowledge represented  as rules in expert systems does  not scale well and
there is no way to  refine the rules and it is also very  difficult to check the coherence
of all  rules within a system.  To summarise, while  the knowledge mined from  data mining
processes  is dynamic  and flexible  but its  representation differs  from one  process to
another. Some knowledge  representation models are rules oriented while  others are stored
as vectors or trained models.

In  this paper,  we propose  a  scalable ontology-based  knowledge map  (OAK) model  for
representing,  storing,  managing  and  retrieving   knowledge  of  any  type.   The  main
contribution of the  model is to support  data scientists and agronomists  in managing and
using mined knowledge for decision making with ease. The next section gives an overview of
knowledge  concepts and  how  to  create knowledge  in  the  agriculture domain.   Section
\ref{sec:OAK} describes in details the proposed OAK model, which includes definitions,
architecture  and  its  three  main   modules.   Section  \ref{sec:Exp}  provides  various
experiments on the  knowledge management system, which is based  on OAK model.  Finally,
we conclude and give some future directions in Section \ref{sec:Concl}.

\section{Related Work}
\label{sec:RW}
According to the Collins
dictionary\footnote{https://www.collinsdictionary.com/dictionary/english/knowledge},
{\it knowledge} is an information and understanding about a subject, which a person has, or
which  all  people have,  while  the  Cambridge dictionary  defines  the  knowledge as  an
information and understanding that you have in your mind.
{\it  Knowledge  Map} (KMap)  is  defined  as  a  spacial representation  of  information
\cite{vail1999knowledge}. Knowledge maps  are used extensively in  enterprises to describe
how,  what  or  where  to  find  useful  knowledge  within  the  enterprise  organisations
\cite{eppler2006comparison}.
{\it Knowledge graph} (KG) is a set of pairs  of knowledge $(V,E)$, where $V$ is a set of
nodes mapped to pieces of knowledge in the  domain and $E$ is a set of knowledge relations
between two nodes.
{\it  Ontology} is  a formal specification  of the vocabulary  to be used  in specifying
knowledge and  the purpose of  the ontology is to  provide a uniform  text-based knowledge
representation,    which    is    comprehensible    by   either    human    or    machines
\cite{kim2003building}.
The difference between these  three concepts (KMap, KG, and ontology)  is that ontology is
used  for formal  and static  knowledge, while  KMaps and  KG are  used for  handling more
dynamic   knowledge    types,   such   as    enterprise   KMaps   or    Google   Knowledge
Graph\footnote{https://developers.google.com/knowledge-graph}. 

In  the following,  discuss some  relevant knowledge  discovery processes  in agriculture,
particularly data mining, and review the knowledge representation and management gaps as a
consequence of proliferation of big data analytics.

\subsection{Knowledge Discovery in Agriculture}
\label{sec:KDDinAgri}

There are many ways  in creating knowledge in any specific domain. The  first method is to
acquire  knowledge manually  by experts.  The result  of this  approach is  knowledge-base
(known as taxonomy  or ontology), or rule-based system (known  as expert systems). Another
way  is  to  acquire  knowledge  automatically via  a  knowledge  discovery  process  from
structured  or  unstructured data.  However,  knowledge  graphs  are normally  built  from
Wikipedia,   Freebase,   DBpedia  and   agriculture   websites,   such  as,   the   AgriKG
\cite{chen2019agrikg} and Cn-MAKG \cite{chenglin2018cn} systems. 

During  the  last  decade, the  advances  in  digital  agriculture  have led  to  numerous
significant studies  on the application of  data mining process to  agricultural datasets.
These  datasets include  including  soil,  weather, crop  yield,  and disease  prediction.
Moreover, the scope of research for data  mining in agriculture is also diverse, including
data  construction as  well as  forecasting  models. In  this context,  there are  several
computational   soil   studies,    such   as   building   datasets    of   soil   profiles
\cite{shangguan2013china}, monitoring soil characteristics  under effects of other factors
and crop yield \cite{bishop2001comparison}, or using soil characteristics to predict other
soil characteristics  \cite{wang2019comparison}. Another  common application  of knowledge
mining in  agriculture is  crop yield  prediction (e.g., predicting  yield or  wheat yield
based    on    soil    attributes,     weather    factors,    and    management    factors
\cite{aggelopoulou2011yield}, \cite{liu2001neural}, \cite{pantazi2016wheat}). Finally, the
third common application  of knowledge mining is the disease  prediction or the protection
plan  (e.g.,  detecting  nitrogen  stress  in  early crop  growth  stage  of  corn  fields
\cite{maltas2013effect},   or    detecting   and    classifying   sugar    beet   diseases
\cite{karimi2006application}, etc.). 

Data  mining  has   four  main  categories  of   techniques;  clustering,  classification,
regression, and association  rule. To date, only classification and  regression are widely
used in  digital agriculture. Usually,  classification approaches  are used to  detect the
disease (mostly  based on the images  processing techniques \cite{karimi2006application}),
or to  classify crop  yield (as  low, medium  or high  yield).  Regression  techniques are
mainly   used   to   predict   crop   yields   based   on   different   input   attributes
\cite{aggelopoulou2011yield},   \cite{bishop2001comparison}.   Although   the  number   of
knowledge model types is not large, input and output attributes for forecasting models are
totally different  and diverse, and the  number of agricultural attributes  is very large.
In addition, the idea  of using KMaps for handling mined knowledge has been proposed    in
\cite{le2007knowledge}   and \cite{le2006admire}, however, it only proposed as a prototype 
for handling rules in   data mining result. Therefore,  the challenge  of these  knowledge 
types and KMaps approach for these knowledge types is that  most of  them are  stored   as
pre-trained models or as computer software and their final results are mostly published as
scientific papers or reports.

\subsection{Methodology for Building Knowledge Map}
\label{sec:methodology4KMap}

There are many approaches for building knowledge maps (KMap).  Most of these methodologies
were dedicated to enterprise KMaps while several  others are used to build specific domain
of KMaps.  For example, Bargent \cite{bargent2002_11steps} proposed an 11-step methodology
for  building an  enterprise KMap,  which  is a  common strategy  of software  development
life-cycle.  Similarly, Kim et al.\cite{kim2003building} proposed a 6-step methodology for
capturing  and representing  organisational  KMaps with  knowledge  profiles and  business
processes, while Pei  et al. \cite{pei2009study} introduced a 7-step  methodology to build
an enterprise  KMap for  matrix organisations,  including setting  up a  development team,
analysis  knowledge resources,  definition of  the business  knowledge domain  boundaries,
determination of the structure and relationship for the KMap, selection of the development
tools, identification  of locations of knowledge  items and drawing the  initial KMap, and
finally evaluation and updates of KMap.

Moreover, several  studies have used  ontology for  building their knowledge  maps. Lecocq
\cite{lecocq2006knowledge}  developed a  4-phase  methodology  with planning,  collecting,
mapping and  validating phases based  on an  ontology for visually  representing knowledge
assets.   In another  study,  Mansingh et  al.   \cite{mansingh2009building} introduced  a
3-stage methodology for building KMaps of medical-care processes; including creation of an
ontology from  the medical cases,  creation of the process  map with flowcharts  and petri
nets,  and  extraction  of  KMaps  from  instances  of  different  medical-care  processes
(represented as medical files) by using vocabulary and relationships in that ontology. The
method was  tested in a healthcare  organisation and was found  to be suitable to  build a
KMap, which combines conceptual and process maps for medical-care processes.

To summarise,  existing methodologies for building  KMaps can be divided  into two typical
types;    methods    for     building    enterprise    KMaps    \cite{bargent2002_11steps}
\cite{kim2003building} \cite{liu2009method}  \cite{pei2009study} and methods  for building
conceptual  KMaps  \cite{lecocq2006knowledge} \cite{mansingh2009building}.   For  building
enterprise KMaps,  although methodologies have different  number of steps, they  use basic
stages:  (1) identify  the  scope, domain  of  KMap  and the  develop  team, (2)  identify
knowledge  resources or  materials, (3)  identify knowledge  in each  knowledge item,  (4)
extract and build KMap,  (5) evaluate, use, and update the  KMap.  For building conceptual
KMaps or  hybrid with  conceptual maps,  existing methodologies also  have the  same basic
stages, however,  one step in  identifying knowledge resources or  materials is to  use an
ontology as a conceptual framework for building KMap in subsequent stages.  This step aims
to locate the knowledge items in the final KMaps.

\section{Ontology-based Knowledge Map Model}
\label{sec:OAK}

\begin{figure}[htbp]
   \centering
   \includegraphics[width=12cm]{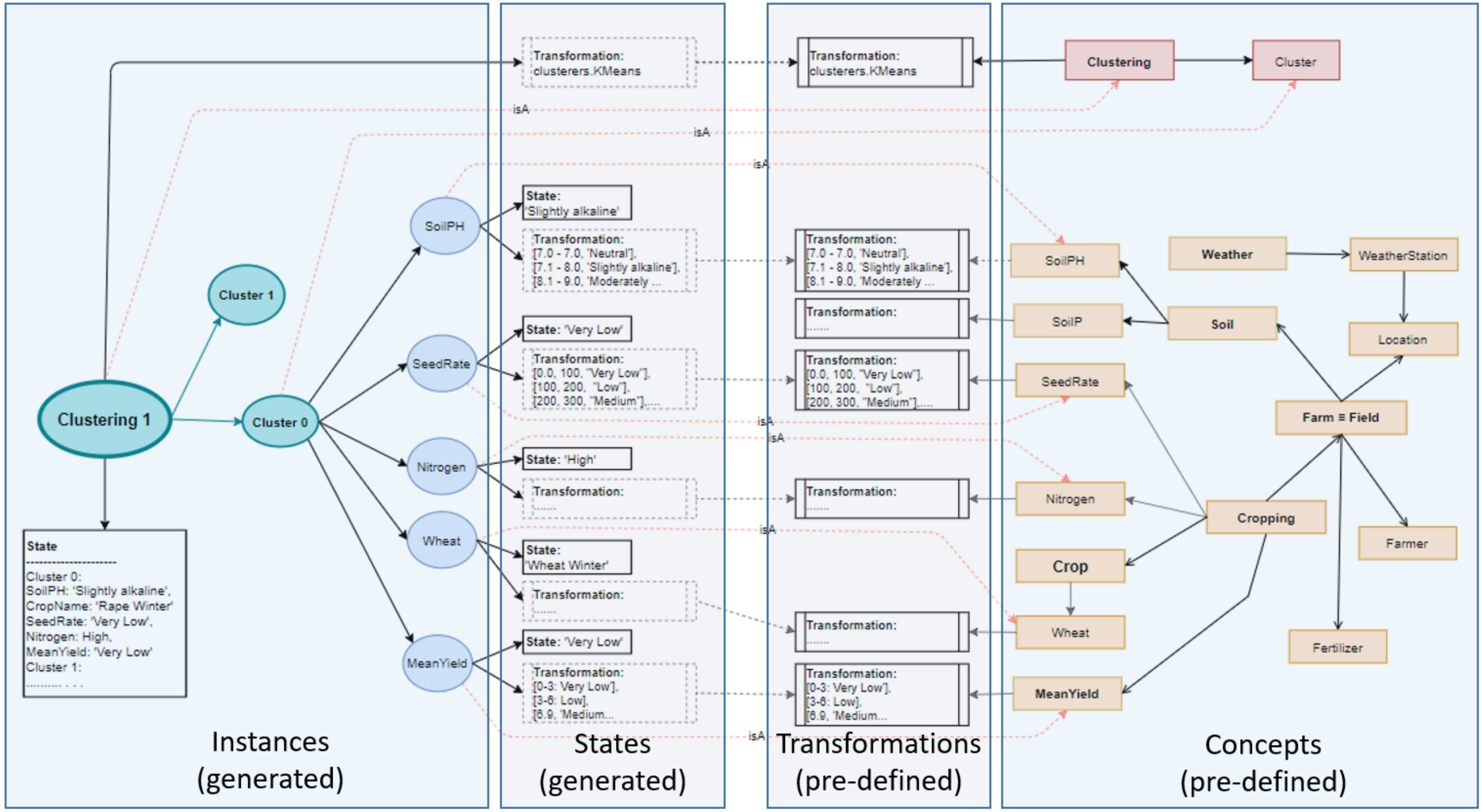}
   \caption{Approach for Knowledge Map Model for Clustering.}
   \label{figCropKMap}
\end{figure}

We propose an  ontology-base knowledge map model for representing  knowledge obtained from
learning processes applied  to agricultural data, from research articles,  or from experts
in  the domain.  Moreover,  the model  allows  knowledge handling  and  exploitation in  a
flexible and scalable  way. As illuminated in an example  in Figure \ref{figCropKMap}, the
knowledge representation shows  the results of a clustering model  with 5 input attributes
(Soil pH, seed rate, nitrogen, wheat name, and mean yield).

To  implement  the model  and  validate  it experimentally,  the  system  is designed  and
developed in two major phases: during the first phase we build an agriculture ontology and
in the second phase  we use the ontology vocabulary along with initial  data schema (as in
the data  warehouse) to model the  knowledge.  Before going  into the details, we  need to
define the key concepts behind this  model, which are {\it knowledge representation}, {\it
  ontology},  {\it  knowledge  map  model},  {\it  concept},  {\it  transformation},  {\it
  instance}, {\it state}, {\it relation}, {\it lexicon} and {\it hierarchy}.

\begin{definition}
  A  Knowledge  representation \(\mathbb{K}\)  is  defined  by  a  set of  four  elements,
  containing    instances   \(\mathbb{I}\),    transformations   \(\mathbb{T}\),    states
  \(\mathbb{S}\), and relations \(\mathbb{R}\):
  \begin{equation}
    \mathbb{K} = (\mathbb{I}, \mathbb{T}, \mathbb{S}, \mathbb{R})
  \end{equation}
\end{definition}

\begin{definition}
  An Ontology  \(\mathbb{O}\) is defined by  a set of three  elements, containing concepts
  \(\mathbb{C}\) , transformations \(\mathbb{T}\), and relations \(\mathbb{R}\):
 \begin{equation}
   \mathbb{O} = (\mathbb{C}, \mathbb{T}, \mathbb{R})
 \end{equation}
\end{definition}

\begin{definition}
  A Knowledge  Map Model  \(\mathbb{KM}\) is  defined by a  set containing  five elements,
  \(\mathbb{C}\),  \(\mathbb{I}\),  \(\mathbb{T}\),  \(\mathbb{S}\),  and  \(\mathbb{R}\),
  which are  corresponding sets  of concept \(c\),  instance \(i\),  transformation \(t\),
  state \(t\) and relation \(r\).
\begin{equation}
	\mathbb{KM} = (\mathbb{C}, \mathbb{I}, \mathbb{T}, \mathbb{S}, \mathbb{R})
\end{equation}
\end{definition}

Where \(\mathbb{C}\) is a  set of concepts \(\{c\}\), and represents a  set of entities or
attributes  of an  entity within  a  domain and  four  data mining  categories of  results
(clustering, classification, regression, and association rule).

\(\mathbb{T}\) is  a set of transformations  \(\{t\}\), and represents a  set of functions
\(f(x)\) to  transform the value  of entities \(x\)  from value range  \(\mathbb{R}_x\) to
value range \(\mathbb{R}_y\).

\(\mathbb{I}\) is a set of instances \(\{i\}\), and represents a set of entities.

\(\mathbb{S}\) is a set of states \(\{s\}\), and represents a set of real attributes of
instances \(\{i\}\) when applying transformation \(\{t\}\).

\(\mathbb{R}\) is a  set of relations \(r\),  and represents a set  of relationships \(r\)
between concept \(c_1\) and concept \(c_2\).

\subsection{Architecture of Knowledge Map Model}
\label{sec:archiKMap}

\begin{figure}[htbp]
 \centering
 \includegraphics[width=12cm]{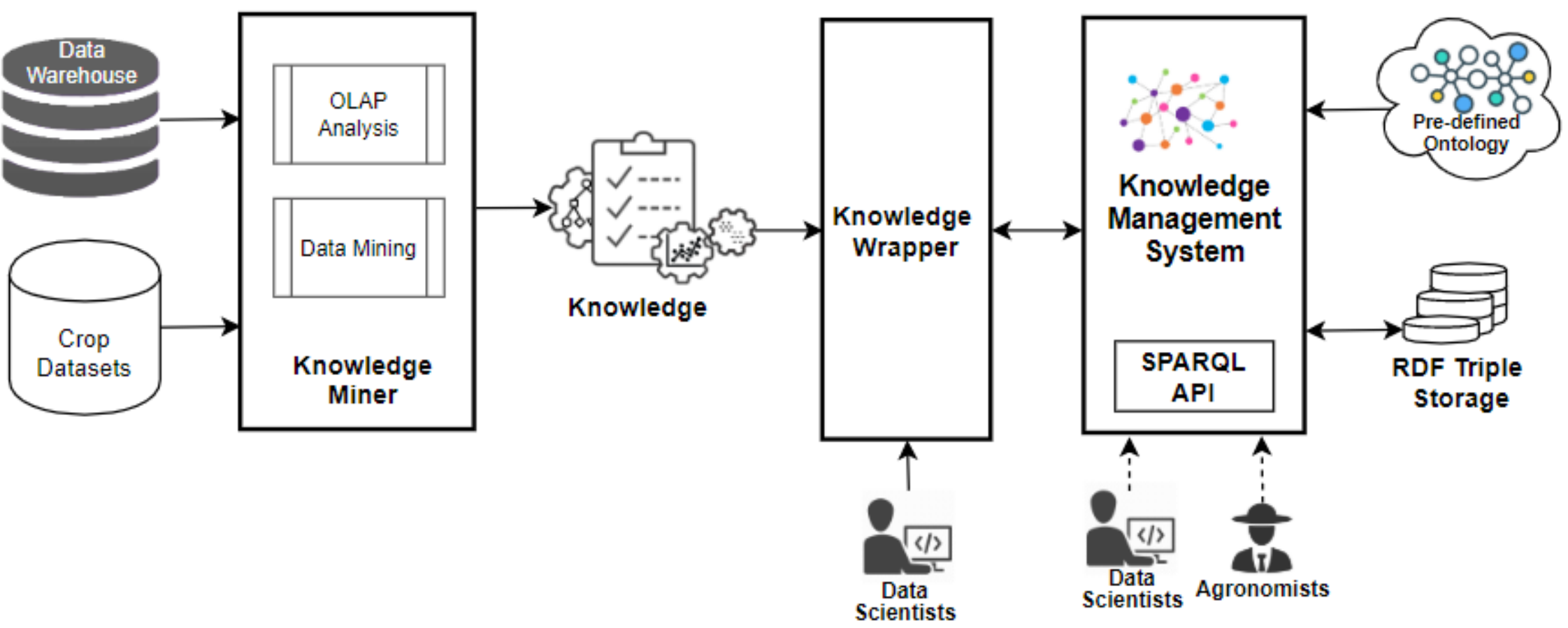}
 \caption{Architecture of Knowledge Map}
 \label{figKMArchitechture}
\end{figure}

As illustrated  in   Figure \ref{figKMArchitechture}, the architecture of  OAK 
consists  of  the following key components: (i) 
Knowledge Miner, (ii)  Knowledge Wrapper and  (iii) Knowledge Management System.
\begin{itemize} 
    \item \textbf{Knowledge Miner} is a key component, which is  used to extract 
	knowledge from data; this component can be a \textit{Data Mining} or an \textit{OLAP Analysis} 
	module.
    \begin{itemize} 
        \item \textbf{Data Mining} refers  to mining tools and techniques, which 
		are used in analyzing datasets from various dimensions and perspectives, 
		finding   hidden    knowledge   and   summarizing   the identified 
		relationships.    These   techniques   are   classification, clustering, 
		regression, association rules.
        \item \textbf{OLAP Analysis}   refers   to   mining   processes  used in 
		analyzing   different   dimensions   of   multidimensional data, which is 
		collected from multiple data sources and stored in data warehouses.
    \end{itemize}
    
    \item {\bf Knowledge Wrapper} is the main module to transform the knowledge
     from the output of the {\it Knowledge Miner} module to the {\it Knowledge 
     Management System} module to store them. This module collects the mining  
     result, identify the type of the data mining task, data mining algorithms, 
     list of agriculture concepts, correlative transformation functions, and 
     states. Then, it generates the KMap representation before converting it into 
     RDF turtles and import into the \textit{Knowledge Management System}.
    
    \item \textbf{Knowledge Management System} is a graph database server, which 
    supports RDF triple storage and SPARQL protocol for the queries. This module
    receives the domain     knowledge   from the pre-defined ontology, the mined 
    knowledge   representations  from the \textit{Knowledge Wrapper} module, and 
    then store in the RDF Triple Storage as a set of RDF turtles. 
\end{itemize} 

In general, the proposed OAK model includes two  knowledge layers.  The first 
layer is the background knowledge  about agriculture.  This layer is defined  as  
a core MKap and it is built from a pre-defined  agricultural ontology (mainly 
cropping    knowledge     in this project).  This layer  defines most of concepts 
(agricultural   entities related to crop)   in     the KMaps and common relations 
between them. The second layer includes knowledge representations  of data mining 
results, which are  mined  from cropping    datasets    by data mining algorithms 
(included in the Knowledge Miner module).   These knowledge representations are 
integrated into the Knowledge Management System by the Knowledge Wrapper module.

\subsection{Agriculture Ontology}
\label{sec:AgriOntology}

In general,   most ontologies describe classes (concepts), instances, attributes, 
and   relations.   Moreover,   some ontologies also include restrictions, rules, 
axioms, and function terms. In our case, as a formal presentation of KMaps, 
we propose an ontology with the following components:

\begin{itemize} 
    \item \textbf{Concepts}:  Concepts   in   the  ontology include concepts in 
        agriculture and concepts for representing four main tasks of data mining.
        For example, agriculture concepts have field, farmer, crop, organization,   
        location,  product, while data      mining   concepts    have clustering, 
        classification,   regression,    and association rule.
    \item \textbf{Transformations}: They are pre-defined transformation functions 
        of agriculture concepts and existing data mining techniques for four main 
        tasks of data mining.
    \item \textbf{Relations}:   ways  in which concepts  (and then instances) can 
        be    related to others.   They are defined  as the \(\mathbb{R}\) set in 
		definitions. 
\end{itemize} 

\begin{figure}[ht]
 \centering
 \includegraphics[width=12cm]{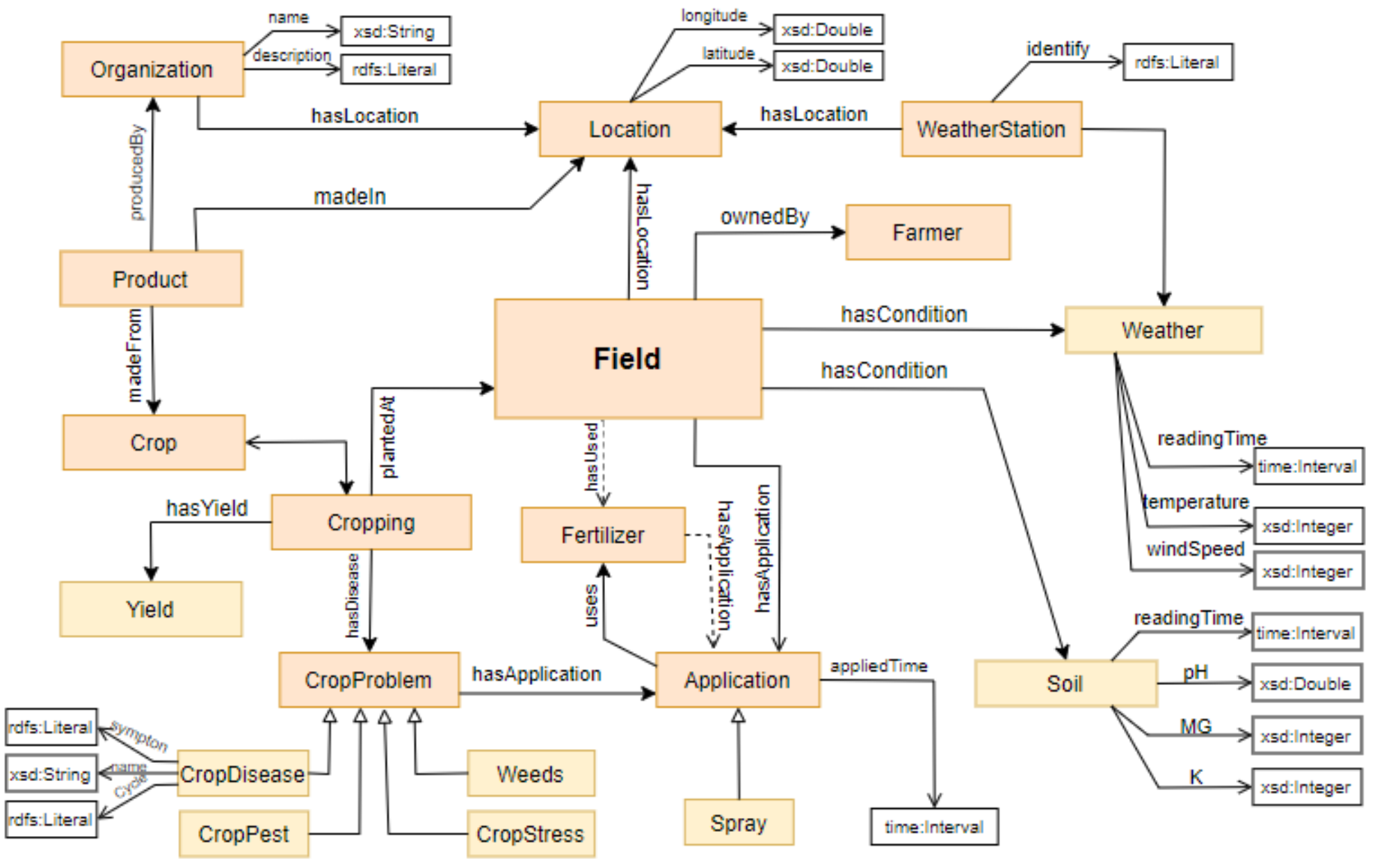}
 \caption{An overview of Agricultural Ontology architecture}
  \label{figAgriOntology}
\end{figure}

\begin{figure}[ht]
 \centering
 \includegraphics[width=12cm]{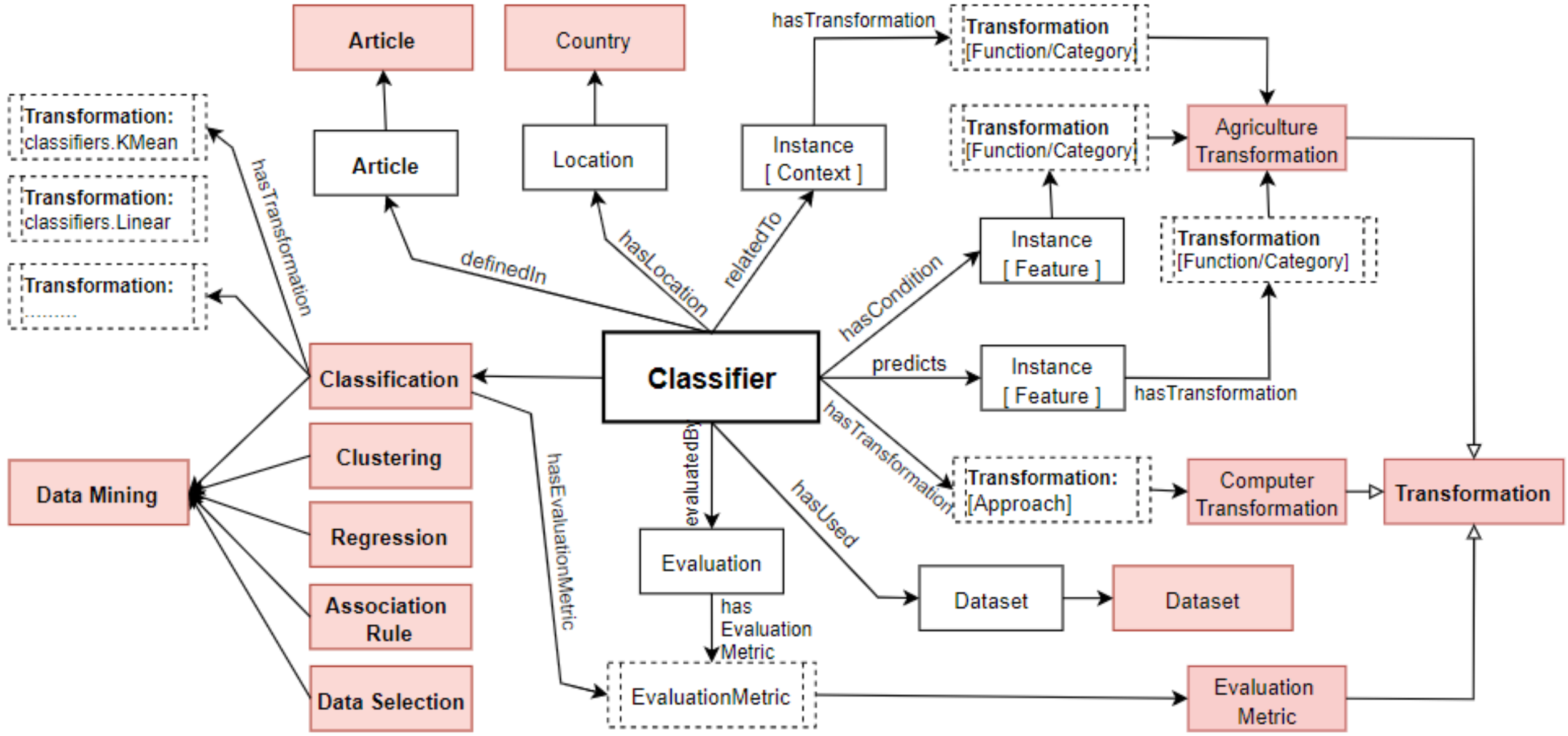}
 \caption{Main data mining concepts in the ontology.}
 \label{figDMconceptInOKM}
\end{figure}

At this stage,  we propose an agricultural ontology AgriOnt for the purpose of 
using it in the OAK model. The agricultural ontology contains 4 sub-domains:  
agriculture, Internet-of-Thing (IoT),  geographical, and the business sub-domain  
(Figure \ref{figAgriOntology}). In addition, the ontology is also added concepts
in data mining domain as shown in Figure \ref{figDMconceptInOKM}. These concepts
and   relations  will be  knowledge   frameworks  to  transform mined  knowledge
from data mining to knowledge representations and import them into the knowledge
maps.

\begin{figure}[ht]
 \centering
 \includegraphics[width=12cm]{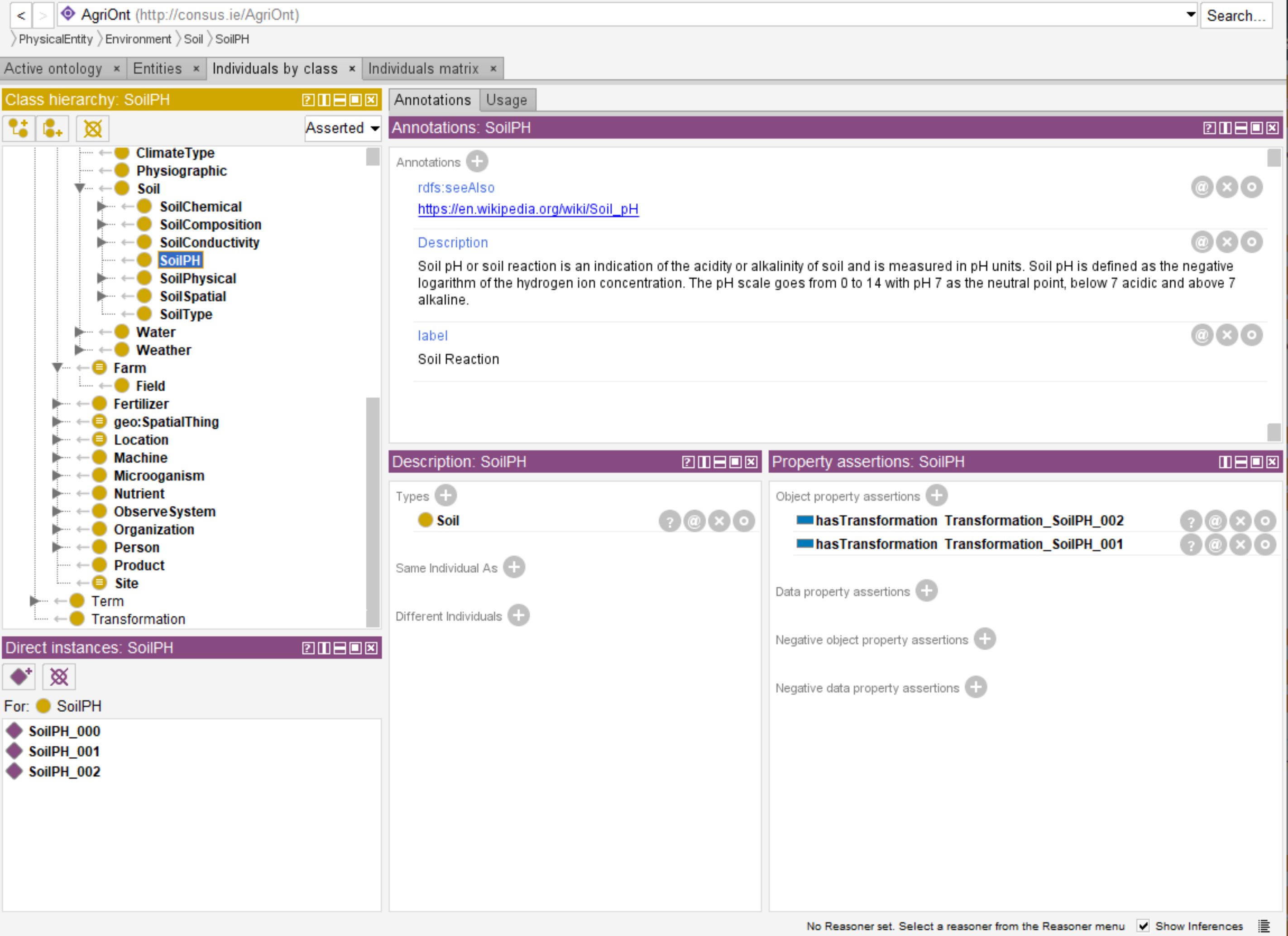}
 \caption{A screenshot of Agricultural Ontology on Protege}
  \label{figureAgriOntOnProtege}
\end{figure}

\begin{table}[ht]
\centering
  \caption{AgriOnt's ontology metrics}
  \label{tableOntologyMetrics}%
    \begin{tabular} {lrrr}
    \hline\noalign{\smallskip}
        \textbf{Figure} & \hspace{0.5cm}\textbf{Core} & 
        \hspace{0.2cm}\textbf{with Transformations} & 
        \hspace{0.2cm}\textbf{with Geo-data}   \\
    \hline\noalign{\smallskip}
        \textbf{Axiom}  	                & 7,947	 &10,484 & 13,917 \\
        \textbf{Logical axiom count} 	    & 3,782  & 5,194 &  7,892 \\
        \textbf{Declaration axioms}         & 1,796  & 2,218 &  2,460 \\
        \textbf{Class count} 	            &   361  &   361 &    361 \\
        \textbf{Object property count}      &    90  &    90 &     90 \\
        \textbf{Data property count}        &   156  &   156 &    156 \\
        \textbf{Individual count}           & 1,183  & 1,605 &  1,847 \\
    \hline\noalign{\smallskip}
    \end{tabular}%
\end{table}%

After building a knowledge hierarchy, the ontology not only provides an overview 
of the agriculture domain  but also describes agricultural concepts,    and life 
cycles between seeds,  plants, harvesting, transportation,  and consumption.  It 
also gives the relationships between agricultural concepts and related  concepts,
such as weather, soil  conditions,  fertilizers, farm descriptions. In addition, 
this    ontology   also   includes data mining concepts, such as classification, 
clustering, regression,     and association rule.   These concepts combined with 
agricultural concepts that are used  to  represent    mined  knowledge. In fact, 
this ontology has 361 classes and over 7,947 axioms   related    to  agriculture
(as      shown in      Table \ref{tableOntologyMetrics}, and partly presented in
\cite{ngo2018ontology}).    As result, the AgriOnt ontology can be  used  as the
core ontology  to    build the knowledge maps for  agriculture. Moreover,   this
ontology with  agricultural  hierarchy can help to integrate available resources
to build larger and more precise knowledge maps in agriculture domain.

\subsection{Knowledge Management System}
\label{sec:KMSystem}

The knowledge consumption is handled in the Knowledge Wrapper module. This 
wrapper transforms   the mined knowledge to the Knowledge  Maps layer. The 
Knowledge Maps layer stores and indexes RDF-based data. The Knowledge Maps layer   
provides   the   Data Access interfaces for query processing SPARQL engine.  The 
SPARQL engine enables the application developers to query data   via    a SPARQL 
Endpoint or a SPARQL-based application  in   the Application layer respectively. 
The SPARQL Endpoint    serves one-shot queries using an extension  of SPARQL 1.1 
query language.

In our approach, we use \textbf{Apache Jena}\footnote{https://jena.apache.org/index.html}
for SPARQL Engine and \textbf{Fuseki}\footnote{https://jena.apache.org/documentation/fuseki2/} 
for SPARQL Endpoint. Both Apache Jena and Fuseki are free and open source. Fuseki
is an HTTP interface to RDF data and it supports SPARQL for querying and updating
data.

\subsection{Knowledge Wrapper}
\label{sec:buildPhase2}

\begin{figure}[ht]
    \centering
    \includegraphics[width=12cm]{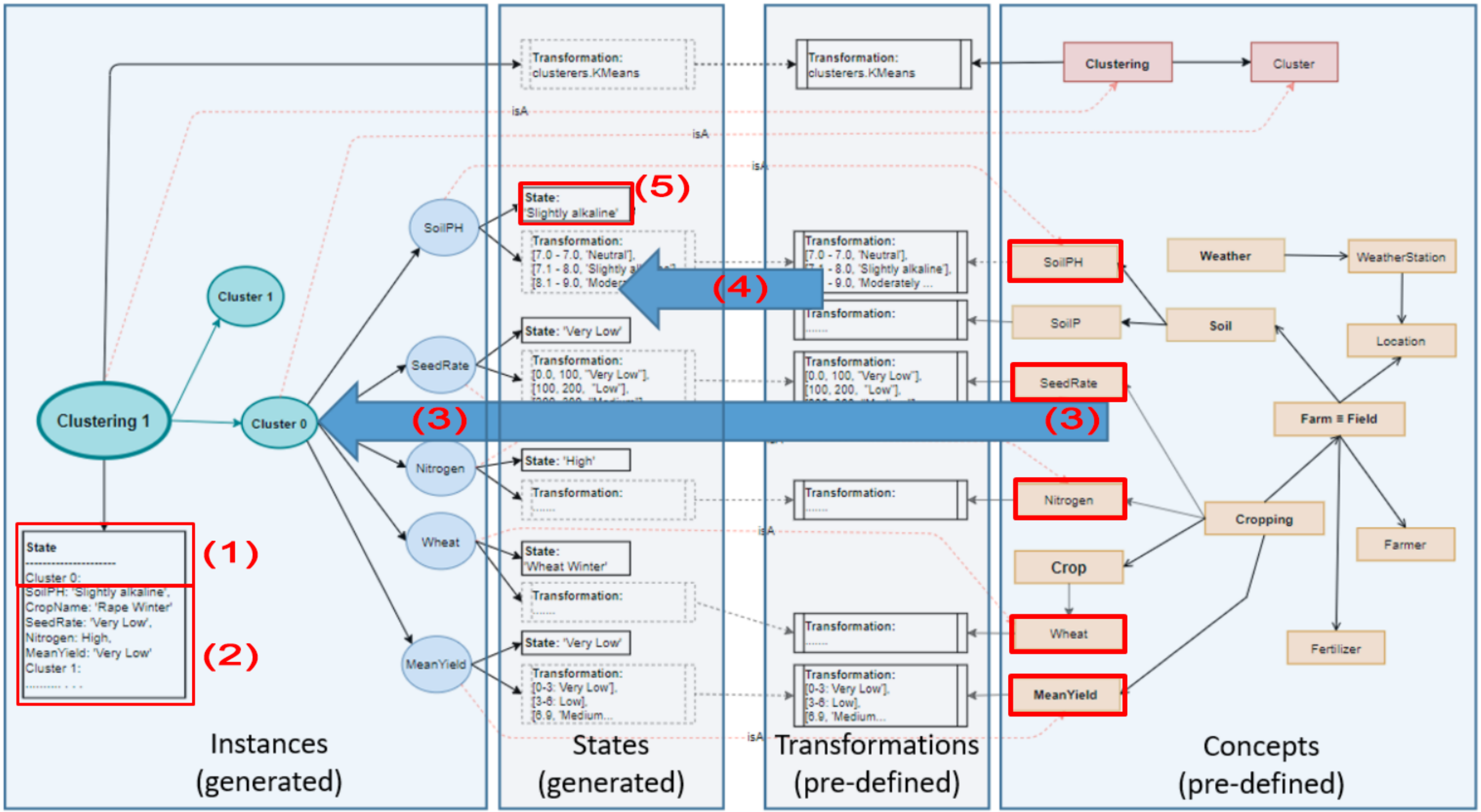}
    \caption{Generating knowledge representation  the Ontology.}
    \label{figOKMapBuildingSteps}
\end{figure}

The procedure for mapping mined knowledge into a knowledge representation in the 
ontology-based knowledge map is defined in the \textit{Knowledge Wrapper} module.
The \textit{Knowledge Wrapper}  module is the main module to transform the mined  
knowledge into a knowledge     representation \(k\) (as defined by \(k = (\{i\}, 
\{t\}, \{s\}, \{r\})\)      in \textit{Definition 1},     Section \ref{sec:OAK})  
before converting into RDF tubles and import them in to the RDF Triple storage.
This module has six steps (Figure \ref{figOKMapBuildingSteps}):
\begin{itemize}
    \item \textbf{Step 1, Identify model}: Select data mining pattern based   on
        the data mining algorithm and generate the data mining instances   (such
        as classification, clustering,    clustering,     and   association rule
        instances  for the corresponding data mining tasks) and link to the data
        mining  algorithm   as    the  transformation objects of the data mining
        instances.
    \item \textbf{Step 2, Identify concepts}: Identify agricultural concepts  in
        mining   results   and locate   them in the ontology. Basically,   these
        concepts occur in the   mined results  as input features and  predicting
        features,   for example, SoilPH, SeedRate, Nitrogen, Wheat, MeanYield.
    \item \textbf{Step 3, Generate instances}: Generate agricultural   instances
        of each located      concept and link them to  data mining instances (in
        \textbf{Step   1}) based on the framework of data mining tasks.
    \item \textbf{Step 4, Identify transformations}:  Identify   transformations
        of each  concept in the mining results, locate them in the ontology part
        of the  KMap, then link them to the agricultural instances (in {\bf Step
        3}). 
    \item \textbf{Step 5, Generate states}: Identify   states  of   each concept
        in the mining   results,  generate  states  and   link  to instances (in
        \textbf{Step 3}) in   the  knowledge representation.
    \item \textbf{Step 6, Generate turtles}:    Transform   the        knowledge
        representation  into  RDF  turtles and import them into the RDF   triple
        storage.
\end{itemize}

In fact, the set of instances \(\{i\}\) is created in  Step 1 and Step 3,  while 
the set of transformations \(\{i\}\) is created in Step 1 and Step 4. The set of 
states      \(\{s\}\)    is generated in   Step 5,    however, not all knowledge
representations have sets of states.  For example, in the model to predict  crop
yield,  the input values only occur when the model is executed.  Therefore,  the
set  of states for this knowledge representation is nearly none. Finally, set of
relations       \(\{i\}\)       is      based   on       relation  \textit{isA},
\textit{hasTransformation},     \textit{hasState},   \textit{hasCondition},  and
\textit{predicts}.

\section{Experimental Results}
\label{sec:Exp}

Firstly,    we have set  up   a SPARQL Server as a knowledge  management  system 
(cf. Section \ref{sec:KMSystem}) with  Apache  Jena  and Fuseki library. The system 
has also created  a new knowledge map with a pre-defined ontology  (as described 
in    Section   \ref{sec:AgriOntology})      and several  samples of   knowledge
representations,  which are created manually by Protege tool  (as shown in Figure
\ref{figureAgriOntOnProtege}).      The SPARQL Endpoint      can  be accessed at
\textit{http://localhost:3030/manage.html} on local machine for query. 

\begin{figure}[ht]
    \centering
    \includegraphics[width=12cm]{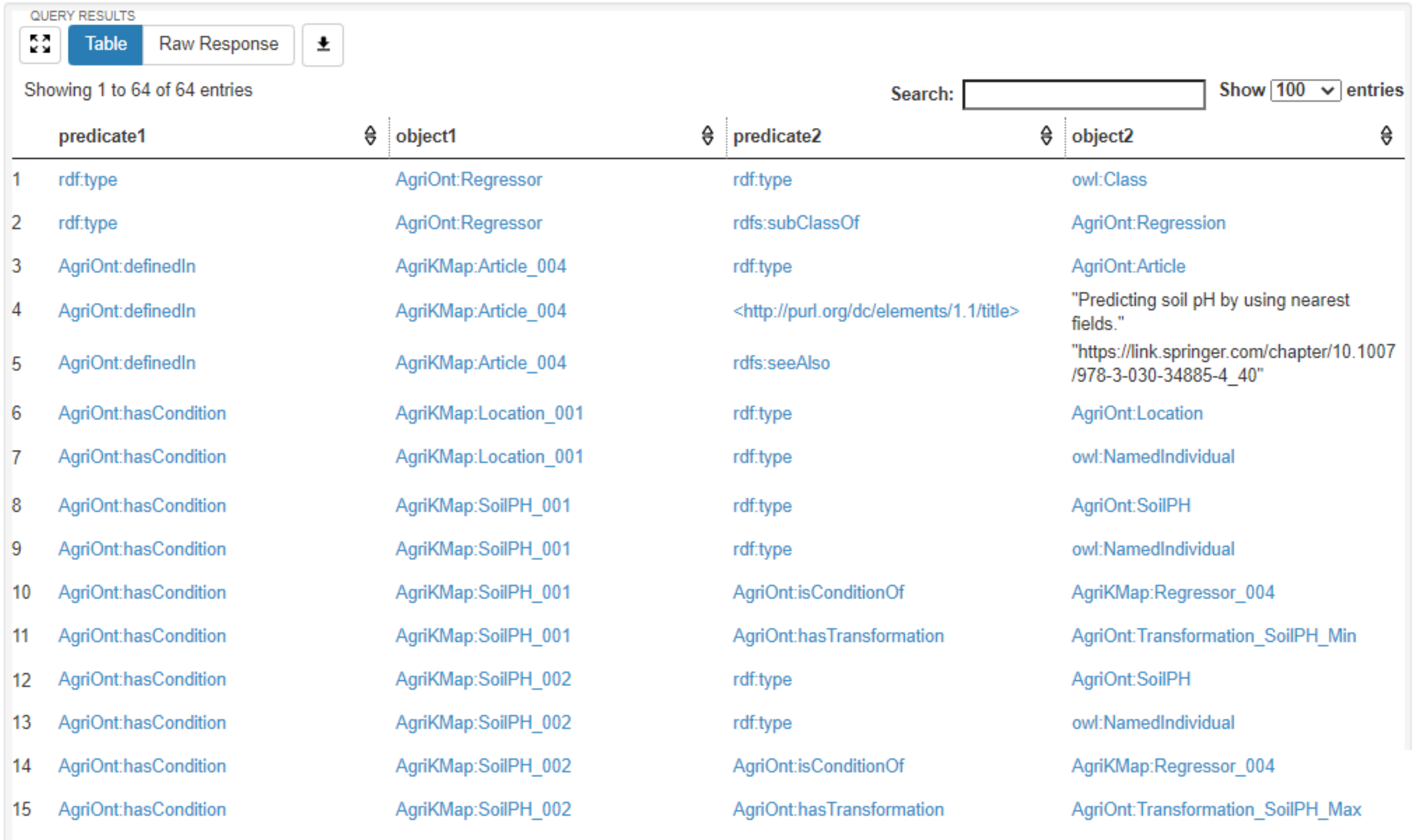}
    \caption{Example of SPARQL query on AgriKMaps.}
    \label{figSPARQLsample}
\end{figure}

Basically, SPARQL queries can be run on the Endpoint with a web interface. Figure
\ref{figSPARQLsample} shows the results for the   query "What are conditions and
the target attribute of the knowledge model \(Regressor\_004\)?".   The   SPARQL
query for this question is shown bellow:

\begin{verbatim}
    PREFIX AgriOnt:  <http://www.ucd.ie/consus/AgriOnt#> 
    PREFIX AgriKMap: <http://www.ucd.ie/consus/AgriKMap#> 
    SELECT ?predicate1 ?object1 ?predicate2 ?object2
    WHERE {
        AgriKMap:Regressor_004 ?predicate1 ?object1 .
        ?object1  ?predicate2 ?object2 .
    }
\end{verbatim}

In   this example,    the     mined knowledge     for         predicting soil pH
\cite{ngo2019predicting}  can    be represented  as a knowledge   representation
\(Regressor\_004\) in the system. Inputs of the model are   pH information   and 
the output of the model is also the pH value (concept \textit{SoilPH}), however,
they used different transformations. Specifically, input  features are calculated
as \(pH_{min}\), \(pH_{max}\), and \(pH_{avg}\) (three different transformations
of SoilPH in the system) of nearest neighbour fields, while the output   feature
is predicted as the original pH value of the field (these conditions are defined
as Soil\_001, Soil\_002, Soil\_003   and Soil\_000 instance, as  shown in Figure
\ref{figSPARQLsample}).

In addition,  the knowledge management system based on the OAK    model can be
benefit  for  both   data scientists  and  agronomists. Data scientists can have
queries about  potential  transformations of agriculture attribute \(c_x\)  (for
example, \textit{Soil pH} or \textit{Temperature}), and these queries  can    be
implemented by a SPARQL query as below:

\begin{verbatim}
    PREFIX AgriOnt:  <http://www.ucd.ie/consus/AgriOnt#> 
    PREFIX AgriKMap: <http://www.ucd.ie/consus/AgriKMap#> 
    SELECT ?subject ?predicate ?object
    WHERE {
        ?subject AgriOnt:transformationOf AgriOnt:SoilPH .
        ?subject ?predicate ?object
    }
\end{verbatim}

For  agronomists   and  also farmers,     they can have queries about potential 
knowledge models or attributes to predict an attribute,   such   as \textit{crop 
yield}. The SPARQL query returns all  potential knowledge  models that are used 
to predict \textit{CropYield} as below:

\begin{verbatim}
    PREFIX rdf: <http://www.w3.org/1999/02/22-rdf-syntax-ns#> 
    PREFIX AgriOnt:  <http://www.ucd.ie/consus/AgriOnt#> 
    PREFIX AgriKMap: <http://www.ucd.ie/consus/AgriKMap#> 
    SELECT ?subject ?predicate ?object
    WHERE {
        ?subject ?predicate ?object .
        ?subject AgriOnt:predicts ?object2 .
        ?object2  rdf:type AgriOnt:CropYield .
    }
\end{verbatim}

\begin{figure}[ht]
    \centering
    \includegraphics[width=12cm]{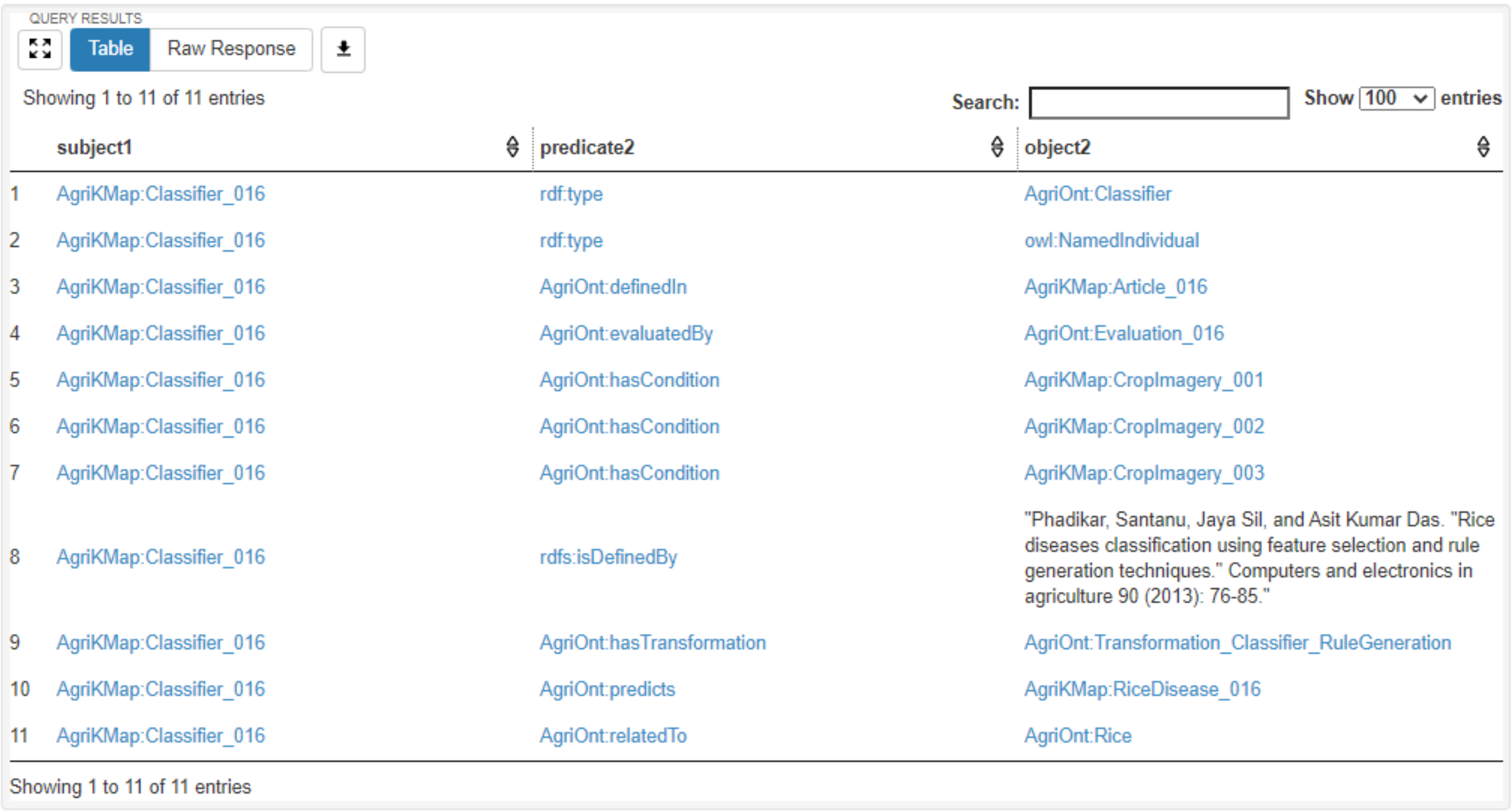}
    \caption{Returns of SPARQL query for Sheath rot disease}
    \label{figSPARQL4StateSample}
\end{figure}

Similarly,  agronomists     also    can have queries for a specific state during 
farming,    such as how to get a \textit{high crop yield} or how to identify the 
{\it  Leaf brown spot disease}. For example, the below query returns a knowledge 
representation   \textit{Classifier\_016},  which   is study of Santanu Phadikar 
\cite{phadikar2013rice}   for   detecting 4 rices diseases, including Leaf brown 
spot, Rice blast, Sheath rot, and Bacterial blight. In which, Sheath rot  is one 
of four predicting  states of the model. The query also provides all information 
related to input attributes,    prediction method   and   evaluation information 
(Figure \ref{figSPARQL4StateSample}).

\begin{verbatim}
    PREFIX rdf: <http://www.w3.org/1999/02/22-rdf-syntax-ns#> 
    PREFIX AgriOnt:  <http://www.ucd.ie/consus/AgriOnt#> 
    PREFIX AgriKMap: <http://www.ucd.ie/consus/AgriKMap#> 
    SELECT ?subject1 ?predicate2 ?object2
    WHERE {
        ?subject1  ?predicate1 ?object1 .
        ?object1   AgriOnt:hasState   AgriOnt:SheathRot .
   		?subject1  ?predicate2 ?object2 .
    }
    LIMIT 100
\end{verbatim}

\section{Conclusion}
\label{sec:Concl}

In this paper,  we present an architecture  for  the OAK - an ontology-based 
knowledge map model  to  represent mined knowledge   from   data mining tasks in
digital  agriculture.   The architecture  includes    Knowledge Miner, Knowledge
Wrapper, Knowledge Management modules based on  an pre-defined ontology. We have
also built an agricultural ontology to provide the domain knowledge  in  agriculture
and a knowledge management system      to  store knowledge representations   and
support the knowledge retrieval efficiently.

With  the proposed ontology-based knowledge map model,  the knowledge management
system  based    on    this model is a promised architecture for handling mined
knowledge in agriculture as well as other domains. As result,  we plan to import
more  knowledge items in    the digital agriculture domain into the    knowledge
management system for retrieval.  Moreover, the knowledge management system also
supports a knowledge browser function as a further method to access knowledge for
both data scientists and agronomists.

\paragraph{\textbf{Acknowledgment}} This work  is part of CONSUS and is supported
by the the SFI Strategic Partnerships Programme    (16/SPP/3296) and is co-funded
by Origin Enterprises Plc.


\begin{thebibliography}{1}

\bibitem{aggelopoulou2011yield}
Aggelopoulou, A.D., Bochtis, D., Fountas, S., Swain, K.C., Gemtos, T.A., Nanos,
G.D.: Yield prediction in apple orchards based on image processing. Precis. Agric.
12(3), 448–456 (2011)

\bibitem{bargent2002_11steps}
Bargent, J.: 11 Steps to Building a Knowledge Map (2002). www.providersedge.
com/docs/km articles/11 Steps to Building a K Map.pdf

\bibitem{bishop2001comparison}
Bishop, T.F.A., McBratney, A.B.: A comparison of prediction methods for the
creation of field-extent soil property maps. Geoderma 103(1–2), 149–160 (2001)

\bibitem{chen2019agrikg}
Chen, Y., Kuang, J., Cheng, D., Zheng, J., Gao, M., Zhou, A.: AgriKG: an agricultural knowledge graph and its applications. In: Li, G., Yang, J., Gama, J.,
Natwichai, J., Tong, Y. (eds.) DASFAA 2019. LNCS, vol. 11448, pp. 533–537.
Springer, Cham (2019). https://doi.org/10.1007/978-3-030-18590-9 81

\bibitem{chenglin2018cn}
Chenglin, Q., Qing, S., Pengzhou, Z., Hui, Y.: Cn-MAKG: China meteorology and
agriculture knowledge graph construction based on semi-structured data. In: 2018
IEEE/ACIS 17th International Conference on Computer and Information Science
(ICIS), pp. 692–696. IEEE (2018)

\bibitem{eppler2006comparison}
Eppler, M.J.: A comparison between concept maps, mind maps, conceptual diagrams, and visual metaphors as complementary tools for knowledge construction
and sharing. Inf. Vis. 5(3), 202–210 (2006)

\bibitem{karimi2006application}
Karimi, Y., Prasher, S.O., Patel, R.M., Kim, S.H.: Application of support vector machine technology for weed and nitrogen stress detection in corn. Comput.
Electron. Agric. 51(1–2), 99–109 (2006)

\bibitem{kim2003building}
Kim, S., Suh, E., Hwang, H.: Building the knowledge map: an industrial case study.
J. Knowl. Manag. 7(2), 34–45 (2003)

\bibitem{le2007knowledge}
Le-Khac, N.-A., Aouad, L.M., Kechadi, M.-T.: Knowledge map: toward a new
approach supporting the knowledge management in distributed data mining. In:
Third International Conference on Autonomic and Autonomous Systems (ICAS
2007), p. 67. IEEE (2007)

\bibitem{le2006admire}
Le-Khac, N.-A., Kechadi, M.T.: Admire framework: distributed data mining on
data-grid platforms. In: International Conference on Software and Data Technologies (ICSOFT 2006), Setubal, Portugal (2006)


\bibitem{lecocq2006knowledge}
Lecocq, R., Valcartier, D.: Knowledge mapping: a conceptual model. Quebec:
Defense Research and Development Canada-Valcartier. Acedido Dezembro, vol.
21, p. 2011 (2006)

\bibitem{liu2001neural}
Liu, J., Goering, C.E., Tian, L.: A neural network for setting target corn yields.
Trans. ASAE 44(3), 705 (2001)


\bibitem{liu2009method}
Liu, L., Li, J., Lv, C.: A method for enterprise knowledge map construction based
on social classification. Syst. Res. Behav. Sci. Off. J. Int. Fed. Syst. Res. 26(2),
143–153 (2009)

\bibitem{maltas2013effect}
Maltas, A., Charles, R., Jeangros, B., Sinaj, S.: Effect of organic fertilizers and
reduced-tillage on soil properties, crop nitrogen response and crop yield: results of
a 12-year experiment in Changins, Switzerland. Soil Tillage Res. 126, 11–18 (2013)


\bibitem{mansingh2009building}
Mansingh, G., Osei-Bryson, K.-M., Reichgelt, H.: Building ontology-based
knowledge maps to assist knowledge process outsourcing decisions. Knowl. Manag.
Res. Pract. 7(1), 37–51 (2009)


\bibitem{ngo2018ontology}
Ngo, Q.H., Le-Khac, N.-A., Kechadi, T.: Ontology based approach for precision
agriculture. In: Kaenampornpan, M., Malaka, R., Nguyen, D.D., Schwind, N. (eds.)
MIWAI 2018. LNCS (LNAI), vol. 11248, pp. 175–186. Springer, Cham (2018).
https://doi.org/10.1007/978-3-030-03014-8 15



\bibitem{ngo2019predicting}
Ngo, Q.H., Le-Khac, N.-A., Kechadi, T.: Predicting soil pH by using nearest fields.
In: Bramer, M., Petridis, M. (eds.) SGAI 2019. LNCS (LNAI), vol. 11927, pp.
480–486. Springer, Cham (2019). https://doi.org/10.1007/978-3-030-34885-4 40


\bibitem{pantazi2016wheat}
Pantazi, X.E., Moshou, D., Alexandridis, T., Whetton, R.L., Mouazen, A.M.:
Wheat yield prediction using machine learning and advanced sensing techniques.
Comput. Electron. Agric. 121, 57–65 (2016)


\bibitem{pei2009study}
Pei, X., Wang, C.: A study on the construction of knowledge map in matrix organizations. In: 2009 International Conference on Management and Service Science
(2009)

\bibitem{phadikar2013rice}
Das, A.K., Phadikar, S., Sil, J.: Rice diseases classification using feature selection
and rule generation techniques. Comput. Electron. Agric. 90, 76–85 (2013)


\bibitem{shangguan2013china}
Shangguan, W., et al.: A China data set of soil properties for land surface modeling.
J. Adv. Model. Earth Syst. 5(2), 212–224 (2013)


\bibitem{vail1999knowledge}
Vail, E.F.: Knowledge mapping: getting started with knowledge management. Inf.
Syst. Manag. 16, 10–23 (1999)

\bibitem{wang2019comparison}
Wang, F., Yang, S., Yang, W., Yang, X., Jianli, D.: Comparison of machine learning
algorithms for soil salinity predictions in three dryland oases located in Xinjiang
Uyghur Autonomous Region (XJUAR) of China. Eur. J. Remote. Sens. 52(1),
256–276 (2019)

\end{thebibliography}

\end{document}